\documentclass[aps,prl,twocolumn,showpacs,superscriptaddress]{revtex4-1}

\usepackage[final]{graphicx}
\usepackage{amsmath}
\usepackage{verbatim}
\usepackage{color}
\usepackage{ulem}

\usepackage{siunitx}
\usepackage[colorlinks,
linkcolor=blue,
citecolor=blue,
urlcolor=blue]{hyperref}

\begin{document}

\preprint{}

\title{Efficient generation of new orbital angular momentum beams by backward and forward stimulated Raman scattering}

	\author{Q. S. Feng} 
    \email{qingsong.feng@physics.ox.ac.uk}
	\affiliation{Department of Physics, Atomic and Laser Physics sub-Department, University of Oxford, Clarendon Laboratory, Parks Road, Oxford OX1 3PU, United Kingdom}

	\author{R. Aboushelbaya} 
	\affiliation{Department of Physics, Atomic and Laser Physics sub-Department, University of Oxford, Clarendon Laboratory, Parks Road, Oxford OX1 3PU, United Kingdom}
	
	\author{M. W. Mayr} 
	\affiliation{Department of Physics, Atomic and Laser Physics sub-Department, University of Oxford, Clarendon Laboratory, Parks Road, Oxford OX1 3PU, United Kingdom}
	
		\author{W. P. Wang} 
		\affiliation{State Key Laboratory of High Field Laser Physics, Shanghai Institute of Optics and Fine Mechanics, Chinese Academy of Sciences, Shanghai 201800, China}
	
		\author{R. M. G. M. Trines} 
		\affiliation{Central Laser Facility, UKRI-STFC Rutherford Appleton Laboratory, Harwell Campus, Didcot, Oxfordshire OX11 0QX, UK}
		
	\author{B. T. Spiers} 
	\affiliation{Department of Physics, Atomic and Laser Physics sub-Department, University of Oxford, Clarendon Laboratory, Parks Road, Oxford OX1 3PU, United Kingdom}
	
	\author{R. W. Paddock} 
	\affiliation{Department of Physics, Atomic and Laser Physics sub-Department, University of Oxford, Clarendon Laboratory, Parks Road, Oxford OX1 3PU, United Kingdom}
	
	\author{I. Ouatu} 
	\affiliation{Department of Physics, Atomic and Laser Physics sub-Department, University of Oxford, Clarendon Laboratory, Parks Road, Oxford OX1 3PU, United Kingdom}

	\author{R. Timmis} 
	\affiliation{Department of Physics, Atomic and Laser Physics sub-Department, University of Oxford, Clarendon Laboratory, Parks Road, Oxford OX1 3PU, United Kingdom}
	
	\author{R. H. W. Wang} 
	\affiliation{Department of Physics, Atomic and Laser Physics sub-Department, University of Oxford, Clarendon Laboratory, Parks Road, Oxford OX1 3PU, United Kingdom}
	
	\author{R. Bingham} 
	\affiliation{Central Laser Facility, UKRI-STFC Rutherford Appleton Laboratory, Harwell Campus, Didcot, Oxfordshire OX11 0QX, UK}

	\author{P. A. Norreys} 
	\affiliation{Department of Physics, Atomic and Laser Physics sub-Department, University of Oxford, Clarendon Laboratory, Parks Road, Oxford OX1 3PU, United Kingdom}
	\affiliation{John Adams Institute, Denys Wilkinson Building, Oxford OX1 3RH, United Kingdom}

\date{\today}

\begin{abstract}
Laser beams carrying orbital angular momentum (OAM) provide an additional degree of freedom and have found wide applications ranging from optical communications and optical manipulation to quantum information. The efficient generation and operation of ultra-intense OAM beams is a big challenge that has to be met, currently setting a limit to the potential applications of ultra-intense OAM beams in high-energy-density physics studies. Here, we theoretically and numerically demonstrate for the first time that a pump beam with a new OAM state is generated by coupling of the seed pulse with OAM Langmuir waves arising from both backward and forward stimulated Raman scattering mechanisms. Advantage is taken of the high energy transfer efficiency from pump to amplified seed beams by operating in the non-linear regime, as this significantly reduces the size of amplification system and promotes access to high-intensity OAM laser beams for scientific and industrial applications.

\end{abstract}


\maketitle


In addition to spin, photons can occupy free-space orbital angular momentum (OAM) eigenstates \cite{Allen_1992PRA}. The phase of an OAM-carrying beam varies with the azimuthal angle $\phi$, i.e., $E\sim \text{exp}( il\phi)$, where $l$ is the topological charge of the OAM state. Beams carrying OAM, a new degree of freedom of light, has attracted substantial attention for myriad applications, including optical communications \cite{Molina_2007NP,Wang_2012NP,Nenad_2013Science}, optical manipulation \cite{Grier_2003Nature,Padgett_2011NP,Wang_2019PRL}, imaging and microscopy \cite{Torner_2005OE}, and quantum information \cite{Mair_2001Nature,Leach_2010Science}. Soft X-ray OAM applications enable the direct alteration of atomic states through orbital angular momentum exchange, and methods to study the electronic properties of quantum materials \cite{Lee_2019NP}.

OAM beams are usually generated by introducing azimuthally dependent phase to the initial Gaussian-shaped laser beam profile using optical elements, such as spatial light modulators \cite{Heckenberg_1992OL}, gratings \cite{Biener_2002OL}, spiral phase plates \cite{Sueda_2004OE,Wang_2020PRL}, and q-plates \cite{Marrucci_2006PRL}. Common schemes use nonlinear optical media for high-harmonic generation and emission of extreme ultra-violet (XUV) OAM lasers \cite{Shao_2013PRA,Gariepy_2014PRL,Zhang_2015PRL,Denoeud_2017PRL,Wang_2019NC} or spiral electron beams in free-electron lasers to produce OAM X-rays \cite{Hemsing_2013NP,Hemsing_2012PRL,Rebernik_2017PRX}. 

The damage threshold of solid state optical elements is usually very low, so it is necessary to operate with intensities lower than $\sim10^{12}$ $W/cm^2$ \cite{Strickland_1985OC}. Because plasma do not have this limitation, they can be used as an ideal medium in which to generate and manipulate ultra-intense OAM laser beams \cite{Shvets_1998PRL,Malkin_1999PRL}. The potential applications of ultra-intense OAM laser beams in high-energy-density physics \cite{Vieira_2018PRL,Ramy_2019PRL,Zhu_2016NC,Zhu_2018NJP} and inertial confinement fusion \cite{Glenzer_2007Nature,Glenzer_2010Science,Zhang_2020NP} call for generating and amplifying ultra-intense OAM beams more flexibly and efficiently. This is a big challenge that has to be met.

Amplification of normal Guassian beams by stimulated Raman scattering \cite{Ren_2007NP,Trines_2011NP,Trines_2011PRL} or stimulated Brillouin scattering \cite{Weber_2013PRL,Kirkwood_2018NP,Marques_2019PRX} in plasma is a hot topic.
The research on Raman/Brillouin amplification of OAM beams is in its infancy. Based on the progress of Raman amplification \cite{Trines_2011NP,Michel_2014PRL}, Vieira \textit{et al.} generated and amplified the short duration seed pulses with new OAM to petawatt intensities by backward stimulated Raman scattering in plasma \cite{Vieira_2016NC}.

In this article, we show for the first time from a new theoretical model supported by three-dimensional (3D) \textit{ab initio} particle-in-cell (PIC) simulations that the seed pulse can couple with the OAM Langmuir wave to generate a new OAM pump by both backward and forward stimulated Raman scattering mechanisms. In addition, by operating in the nonlinear regime for both the pump and seed pulses, the energy transfer efficiency is demonstrated to remain above $50\%$ \cite{Sadler_2017PRE}. These results pave a way to generate long duration pump beams with new OAM states by both backward and forward stimulated Raman scattering, as well as multi-stage efficient Raman amplification to generate high-intensity OAM/Gaussian beams \cite{Huang_2020CP,Ren_2007NP}.

From the Maxwell equations, one obtains the laser propagation equation:
\begin{equation}
\label{Eq:LaserPropagation}
(\frac{\partial^2}{\partial t^2}-c^2\nabla^2){\bf A}=-\frac{4\pi e^2}{m_e}n_e{\bf A},
\end{equation}
where ${\bf A}={\bf A}_{pump}+{\bf A}_{seed}$ is the vector potential of the electromagnetic field of the combined pump and seed pulses, $n_e=n_{e0}+\delta n_e$ is the electron density comprising an equilibrium state density $n_{e0}$ and a density fluctuation $\delta n_e$.
By substituting ${\bf A}={\bf A}_{pump}+{\bf A}_{seed}$ and $n_e=n_{e0}+\delta n_e$ into Eq. (\ref{Eq:LaserPropagation}), one obtains
\begin{equation}
\label{Eq:LaserPropagation_pump}
(\frac{\partial^2}{\partial t^2}-c^2\nabla^2+\omega_{pe}^2){\bf A}_{pump}=-\omega_{pe}^2\frac{\delta n_e}{n_{e0}}{\bf A}_{seed},
\end{equation}
\begin{equation}
\label{Eq:LaserPropagation_seed}
(\frac{\partial^2}{\partial t^2}-c^2\nabla^2+\omega_{pe}^2){\bf A}_{seed}=-\omega_{pe}^2\frac{\delta n_e}{n_{e0}}{\bf A}_{pump},
\end{equation}
From the continuity and momentum equations for the electrons, considering background immobile ions, one obtains the electron density perturbation equation, which in Gaussian units reads:
\begin{equation}
\label{Eq:DensityEquation}
(\frac{\partial^2}{\partial t^2}-3v_{te}^2\nabla^2+\omega_{pe}^2)\frac{\delta n_e}{n_{e0}}=\frac{e^2}{2m_e^2c^2}\nabla^2(2{\bf A}_{pump}{\bf A}_{seed}),
\end{equation}
where $\omega_{pe}=\sqrt{4\pi n_{e0}e^2/m_e}$ is the plasma frequency, and $v_{te}$ is the electron thermal velocity.

Considering ${\bf A}_{pump,seed}=\frac{1}{2}{\bf A_{0,s}}exp(i\psi_{0,s})+c.c.$ and $\tilde{\delta n_e}=\delta n_e/n_{e0}=\frac{1}{2}\tilde{\delta n_e}exp(i\psi_{L})+c.c.$ (where $\psi_j=k_jx-\omega_j t$ ($j=0,s,L$) is phase and c.c. refers to complex conjugate), one can obtain the operators $\partial^2/\partial t^2\rightarrow(-\omega^2-i2\omega\partial/\partial t+\partial^2/\partial t^2)$, $\partial^2/\partial x^2\rightarrow(-k^2+i2k\partial/\partial x+\partial^2/\partial x^2)$, and $\nabla_\bot^2\rightarrow\nabla_\bot^2$. Substituting the above operators into Eqs. (\ref{Eq:LaserPropagation_pump})-(\ref{Eq:DensityEquation}), neglecting the second partial derivatives of the (slowly varying) envelope amplitude ${\bf A_{0,s}}$ and $\tilde{\delta n_e}$, and applying the linear dispersion relation $\omega_{0,s}^2=\omega_{pe}^2+k_{0,s}^2c^2$ and $\omega_{L}^2=\omega_{pe}^2+3k_{L}^2v_{te}^2$, one obtains:
\begin{equation}
\label{Eq:LaserPropagation_pump_envelop}
(i2\omega_0\frac{\partial}{\partial t}+i2k_0c^2\frac{\partial}{\partial x}+c^2\nabla_\bot^2){\bf A_0}=\frac{1}{2}\omega_{pe}^2\tilde{\delta n_e}{\bf A_s},
\end{equation}

\begin{equation}
\label{Eq:LaserPropagation_seed_envelop}
(i2\omega_s\frac{\partial}{\partial t}+i2k_sc^2\frac{\partial}{\partial x}+c^2\nabla_\bot^2){\bf A_s}=\frac{1}{2}\omega_{pe}^2\tilde{\delta n_e}^*{\bf A_0},
\end{equation}

\begin{equation}
\label{Eq:Density_envelop}
(i2\omega_L\frac{\partial}{\partial t}+i2k_LS_e^2\frac{\partial}{\partial x}+S_e^2\nabla_\bot^2)\tilde{\delta n_e}=\frac{1}{2}\frac{e^2k_L^2}{m_e^2c^2}{\bf A_0}{\bf A_s^*},
\end{equation}
where $S_e^2=3v_{te}^2$. Noting that $k_s=-|k_s|$ refers to the backward-SRS wave vector, and $k_s=|k_s|$ refers to the forward-SRS wave vector. The phase matching $\psi_0=\psi_s+\psi_L$ is considered in Eqs. (\ref{Eq:LaserPropagation_pump_envelop})-(\ref{Eq:Density_envelop}), i.e. $\omega_0=\omega_s+\omega_L$ and $k_0=k_s+k_L$.

In our simulations, the laser propagates along the longitudinal direction $x$, and transverse directions are the y and z directions. The vector potential envelop ${\bf A_{j}}$ $(j=0,s)$ of each laser consists of a longitudinal part (${\bf A_{\parallel}}$) and a transverse part (${\bf T}$), i.e.
\begin{equation}
\begin{split}
{\bf A_{j}}(t,{\bf r}_{\perp},x)={\bf A_{j,\parallel}}(t,x){\bf T_j}({\bf r}_{\perp},x).
\end{split}
\end{equation}
${\bf A_{\parallel}}$ is independent of transverse position ${\bf r_{\perp}}$ and ${\bf T}$ is independent of time $t$. Similarly, the density fluctuation in $x$ direction can be expressed as $\tilde{\delta n_e}={\delta n_{\parallel}}{T_{\delta n}}$ which is coupled with the pump and seed pulses.

For a Laguerre-Gaussian (LG) beam with OAM number $l$ and radial mode $p$, the transverse envelop profile ${\bf T_j}$ ($j=0,s$) or $T_{\delta n}$ is described as: \cite{Allen_1992PRA,Zhang_2015PRL,Vieira_2016NC}
\begin{equation}
\begin{aligned}
&{\bf T}({\bf r}_{\perp},x)=\frac{w_0}{w(x)}\bigg(\frac{\sqrt{2}r}{w(x)}\bigg)^{|l|}L_p^{|l|}\bigg(\frac{2{\bf r}_{\perp}^2}{w^2(x)}\bigg)\text{exp}\bigg(-\frac{2{\bf r}_{\perp}^2}{w^2(x)}\bigg)\\
&\times \text{exp}\bigg[\frac{ik{\bf r}_{\perp}^2x}{2(x^2+x_R^2)}\bigg]\text{exp}\bigg[-i(2p+|l|+1)\text{arctan}(\frac{x}{x_R})\bigg]
\\
&\times\text{exp}(il\phi),
\end{aligned}
\end{equation}
where ${\bf r_{\perp}}=y{\bf e_y}+z{\bf e_z}$ is the transverse position and $r=\sqrt{(y^2+z^2)}$ the radial distance to the laser axis, $w(x)=w_0\sqrt{1+x^2/x_R^2}$ is the beam waist, $w_0$ the waist at the focal plane, $x_R=\pi w_0^2/\lambda$ is the Raleigh length, $\lambda$ the central wavelength of the laser beam, $L_p^{|l|}$ is an associated Laguerre polynomial, and $(2p+|l|+1)\text{arctan}(x/x_R)$ is the Gouy phase. 

The conditions of $x\ll x_R$ and $p=0$ are satisfied in all of our simulations, thus the transverse envelop profile of the LG beam with OAM number $l$ can be simplified to read:
\begin{equation}
\begin{split}
{\bf T}({\bf r}_{\perp},x)={\bf T}({\bf r}_{\perp})=\bigg(\frac{\sqrt{2}{r}}{w_0}\bigg)^{|l|}\text{exp}\bigg(-\frac{2{\bf r}_{\perp}^2}{w_0^2}\bigg)\text{exp}(il\phi),
\end{split}
\end{equation}
which is independent with time $t$ and longitudinal position $x$.

Since $\partial{{\bf T_j}}/\partial t=0$, $\partial{{\bf T_j}}/\partial x=0$ ($j=0,s,\delta n$) and $\nabla_{\perp}{\bf A_{j,\parallel}}=\nabla_{\perp}{\delta n_{\parallel}}=0$ ($j=0,s$) when $x\ll x_R$ is satisfied, by using the paraxial approximation $(i2k_j\frac{\partial}{\partial x}+\nabla_\bot^2){\bf T_j}=0$ ($j=0,s$) and $(i2k_L\frac{\partial}{\partial x}+\nabla_\bot^2){T_{\delta n}}=0$, the three wave coupling equations (\ref{Eq:LaserPropagation_pump_envelop})-(\ref{Eq:Density_envelop}) simplify to:
\begin{equation}
\label{Eq:LaserPropagation_pump_envelop1}
(i2\omega_0\frac{\partial}{\partial t}+i2k_0c^2\frac{\partial}{\partial x}){\bf A_0}=\frac{1}{2}\omega_{pe}^2\tilde{\delta n_e}{\bf A_s},
\end{equation}
\begin{equation}
\label{Eq:LaserPropagation_seed_envelop1}
(i2\omega_s\frac{\partial}{\partial t}+i2k_sc^2\frac{\partial}{\partial x}){\bf A_s}=\frac{1}{2}\omega_{pe}^2\tilde{\delta n_e}^*{\bf A_0},
\end{equation}
\begin{equation}
\label{Eq:Density_envelop1}
(i2\omega_L\frac{\partial}{\partial t}+i2k_LS_e^2\frac{\partial}{\partial x})\tilde{\delta n_e}=\frac{1}{2}\frac{e^2k_L^2}{m_e^2c^2}{\bf A_0}{\bf A_s^*}.
\end{equation}
Note that ${\bf A_j}$ ($j=0,s$) and $\tilde{\delta n_e}$ in above equations are functions of $t$, $x$ and ${\bf r_{\perp}}$, i.e. ${\bf A_j}(t,{\bf r_{\perp}},x)$ and $\tilde{\delta n_e}(t,{\bf r_{\perp}},x)$.  

In our simulations, $x\ll x_R$, the paraxial approximation is strictly satisfied. It is well established that the three wave coupling equations  (\ref{Eq:LaserPropagation_pump_envelop1})-(\ref{Eq:Density_envelop1}) are independent of the laser transverse profiles, as long as the paraxial approximation is satisfied. From equations (\ref{Eq:LaserPropagation_pump_envelop1})-(\ref{Eq:Density_envelop1}), one is satisfied that not only the energy conservation ($\omega_0=\omega_s+\omega_L$) and linear momentum conservation ($k_0=k_s+k_L$) are satisfied, but also the conserviation of OAM ($l_0=l_s+l_L$).

To derive the growth rate to explore the temporal problem, assuming $k_{j}\partial/\partial x\ll \omega_{j}\partial/\partial t$ ($j=0,s,L$), the spatial term is negligible. Thus, Eqs. (\ref{Eq:LaserPropagation_pump_envelop1})-(\ref{Eq:Density_envelop1}) can be simplified as
\begin{equation}
\label{Eq:LaserPropagation_pump_envelop2}
\frac{\partial {\bf A_0}}{\partial t}=-i\alpha_0\tilde{\delta n_e}{\bf A_s},
\end{equation}
\begin{equation}
\label{Eq:LaserPropagation_seed_envelop2}
\frac{\partial {\bf A_s}}{\partial t}=-i\alpha_s\tilde{\delta n_e}^*{\bf A_0},
\end{equation}
\begin{equation}
\label{Eq:Density_envelop2}
\frac{\partial \tilde{\delta n_e}}{\partial t}=-i\alpha_L{\bf A_0}{\bf A_s^*}. 
\end{equation}
where $\alpha_0=\frac{\omega_{pe}^2}{4\omega_0}$, $\alpha_s=\frac{\omega_{pe}^2}{4\omega_s}$ and $\alpha_L=\frac{1}{4}\frac{e^2k_L^2}{\omega_Lm_e^2c^2}$. The time derivative of Eq. (\ref{Eq:Density_envelop2}) is 
\begin{equation}
\label{Eq:Density_envelop3}
\frac{\partial^2 \tilde{\delta n_e}}{\partial t^2}=-i\alpha_L({\bf A_0}\frac{\partial {\bf A_s^*}}{\partial t}+{\bf A_s^*}\frac{\partial {\bf A_0}}{\partial t}). 
\end{equation}
Substituting Eqs. (\ref{Eq:LaserPropagation_pump_envelop2}) and (\ref{Eq:LaserPropagation_seed_envelop2}) into Eq. (\ref{Eq:Density_envelop3}), one obtains
\begin{equation}
\label{Eq:Density_envelop4}
\frac{\partial^2 \tilde{\delta n_e}}{\partial t^2}=\Gamma^2\tilde{\delta n_e},
\end{equation}
\begin{equation}
\label{Eq:Gamma}
\Gamma^2=\alpha_L\alpha_s|{\bf A_0}|^2-\alpha_L\alpha_0|{\bf A_s}|^2.
\end{equation}
Considering $\tilde{\delta n_e}(t=0)=0$ and $\partial\tilde{\delta n_e}(t=0)/\partial t=-i\alpha_L{\bf A_{00}}{\bf A_{s0}^*}$ (where ${\bf A_{j0}}\equiv{\bf A_{j}}(t=0)$, $j=0,s$),  one obtains
\begin{equation}
\label{Eq:Density}
\tilde{\delta n_e}=-i\frac{\alpha_L}{\Gamma}{\bf A_{00}}{\bf A_{s0}^*}\text{sinh}(\Gamma t).
\end{equation}

It is possible to derive evolutions of pump pulse and seed pulse.
Firstly, combining Eq. (\ref{Eq:Density}) and Eq. (\ref{Eq:LaserPropagation_seed_envelop2}), and considering the initial condition of ${\bf A_s}(t=0)={\bf A_{s0}}$, one obtains the time evolution of the seed pulse:
\begin{equation}
\label{Eq:As}
{\bf A_s}(t,{\bf r_\bot})=({\bf A_{s0}\cdotp\frac{{\bf A_{00}^*}}{|{\bf A_{00}}|}})\frac{{\bf A_0}}{|{\bf A_{00}}|}\text{cosh}(\Gamma t)+{\bf C_1},
\end{equation}
where ${\bf C_1}$ is an integration constant decided by the initial seed vector potential profile which depends on the initial overlap between pump and seed pulses. At early times while the new OAM seed pulse is still lower than that of the pump pulse, the condition $|{\bf A_s}|^2\ll|{\bf A_0}|^2$ is strictly satisifed. As such, the growth rate of the new OAM seed amplification at these early times can be expressed as:
\begin{equation}
\label{Eq:Gamma1}
\Gamma=\sqrt{\alpha_L\alpha_s}|{\bf A_0}|=\frac{1}{4}\sqrt{\frac{\omega_{pe}^2}{\omega_s\omega_L}}k_L\frac{e|{\bf A_0}|}{m_ec}.
\end{equation}

Secondly, combining Eq. (\ref{Eq:Density}) and Eq. (\ref{Eq:LaserPropagation_pump_envelop2}), and considering the initial condition of ${\bf A_0}(t=0)={\bf A_{00}}$, one obtains
\begin{equation}
\label{Eq:A0}
{\bf A_0}(t,{\bf r_\bot})=({\bf A_{00}\cdotp\frac{{\bf A_{s0}^*}}{|{\bf A_{s0}}|}})\frac{{\bf A_s}}{|{\bf A_{s0}}|}\text{cosh}(\Gamma t)+{\bf C_2}.
\end{equation}
When the seed pulse couples with the density fluctuation (or Langmuir wave) to generate the new pump with new OAM, $|{\bf A'_0}|^2\ll|{\bf A_s}|^2$ is strictly satisfied. In the following, ${\bf A'_{00}}$ and ${\bf A'_0}$ are taken as the initial and time-depending vector potentials of the new pump to distinguish the amplitude of this new pump from the original pump. Thus the growth rate of the new pump $\Gamma'$ in Eq. (\ref{Eq:Gamma1}) is $\Gamma'^2=\alpha_L\alpha_s|{\bf A'_0}|^2-\alpha_L\alpha_0|{\bf A_s}|^2\simeq-\alpha_L\alpha_0|{\bf A_s}|^2$. One can obtain
\begin{equation}
\label{Eq:Gamma'}
\Gamma'=i\sqrt{\alpha_L\alpha_0}|{\bf A_s}|=i\gamma.
\end{equation}
Replacing $\Gamma$ in Eq. (\ref{Eq:Density}) and Eq. (\ref{Eq:A0}) by $\Gamma'$ in Eq. (\ref{Eq:Gamma'}), one obtains
\begin{equation}
\label{Eq:Density1}
\tilde{\delta n'_e}=-i\frac{\alpha_L}{\gamma}{\bf A'_{00}}{\bf A_{s0}^*}\text{sin}(\gamma t),
\end{equation}
\begin{equation}
\label{Eq:A0_1}
{\bf A'_0}(t,{\bf r_\bot})=({\bf A'_{00}\cdotp\frac{{\bf A_{s0}^*}}{|{\bf A_{s0}}|}})\frac{{\bf A_s}}{|{\bf A_{s0}}|}\text{cos}(\gamma t)+{\bf C_2},
\end{equation}
\begin{equation}
\label{Eq:gamma}
\gamma=\sqrt{\alpha_L\alpha_0}|{\bf A_s}|=\frac{1}{4}\sqrt{\frac{\omega_{pe}^2}{\omega_0\omega_L}}k_L\frac{e|{\bf A_s}|}{m_ec},
\end{equation}
where ${\bf C_2}$ is an integration constant decided by the initial new pump vector potential profile which depends on the initial overlap between the pump and seed pulses. 

In our simulations, the pump and seed pulses are both circular-polarisation (CP) Laguerre-Gaussian (LG) modes which are expressed as ${\bf A^{(')}_{0(0)}}=A^{(')}_{0(0),y}\text{exp}(il_{0y}\phi){\bf e_y}+iA^{(')}_{0(0),z}\text{exp}(il_{0z}\phi){\bf e_z}$ and  ${\bf A_{s(0)}}=A_{s(0),y}\text{exp}(il_{sy}\phi){\bf e_y}+iA_{s(0),z}\text{exp}(il_{sz}\phi){\bf e_z}$, where $\text{exp}(il\phi)$ is shown in the term of every polarisation direction, ${\bf A^{(')}_{0(0)}}$ refers to ${\bf A_{0}}$, ${\bf A^{(')}_{0}}$, ${\bf A_{00}}$ and ${\bf A^{'}_{00}}$, and ${\bf A_{s(0)}}$ refers to ${\bf A_{s}}$ and ${\bf A_{s0}}$. Substituting the above expressions into Eqs. (\ref{Eq:Density}), (\ref{Eq:As}) and (\ref{Eq:A0_1}), one obtains
\begin{equation}
\begin{aligned}
\label{Eq:Density2}
\tilde{\delta n_e}=&-i\frac{\alpha_L}{\Gamma}\text{sinh}(\Gamma t)\cdot(A_{00,y}A_{s0,y}^*\text{exp}[i(l_{0y}-l_{sy})\phi]
\\&
+A_{00,z}A_{s0,z}^*\text{exp}[i(l_{0z}-l_{sz})\phi]), 
\end{aligned}
\end{equation}
\begin{equation}
\begin{aligned}
\label{Eq:As_1}
{\bf A_s}=
&\text{cosh}(\Gamma t)\cdot(A_{s0,y}\text{exp}[i(l_{sy})\phi]{\bf e_y}
+iA_{s0,z}\text{exp}[i(l_{sz})\phi]{\bf e_z}
\\&
+(A_{s0,z})\text{exp}[i(l_{sz}-l_{0z}+l_{0y})\phi]{\bf e_y}
\\&
+i(A_{s0,y})\text{exp}[i(l_{sy}-l_{0y}+l_{0z})\phi]{\bf e_z})+{\bf C_1},
\end{aligned}
\end{equation}
\begin{equation}
\begin{aligned}
\label{Eq:A0'_1}
{\bf A'_0}=&
\text{cos}(\gamma t)\cdot(A'_{00,y}\frac{A_{sy}}{A_{s0,y}}\text{exp}[i(l_{0y})\phi]{\bf e_y}
\\&
+iA'_{00,z}\frac{A_{sz}}{A_{s0,z}}\text{exp}[i(l_{0z})\phi]{\bf e_z}
\\&
+A'_{00,z}\frac{A_{sy}}{A_{s0,y}}\text{exp}[i(l_{0z}-l_{sz}+l_{sy})\phi]{\bf e_y}
\\&
+iA'_{00,y}\frac{A_{sz}}{A_{s0,z}}\text{exp}[i(l_{0y}-l_{sy}+l_{sz})\phi]{\bf e_z})+{\bf C_2},
\end{aligned}
\end{equation}
where ${\bf |A_j|}=A_{jy}=A_{jz}$ is satisfied in CP laser for both pump ($j=0$) and seed ($j=s$). The above expressions are simplified as:
\begin{equation}
\label{Eq:Density3}
\tilde{\delta n_e}\propto{\bf A_{00}}{\bf A_{s0}^*}\propto(\text{exp}[i(l_{0y}-l_{sy})\phi]+\text{exp}[i(l_{0z}-l_{sz})\phi]), 
\end{equation}
\begin{equation}
\begin{aligned}
\label{Eq:As_2}
{\bf A_s}\propto
&({\bf A_{s0}\cdotp\frac{{\bf A_{00}^*}}{|{\bf A_{00}}|}})\frac{{\bf A_0}}{|{\bf A_{00}}|}\propto
\\&
(\text{exp}[i(l_{sy})\phi]{\bf e_y}+\text{exp}[i(l_{sz}-l_{0z}+l_{0y})\phi]{\bf e_y}+
\\&
i\text{exp}[i(l_{sz})\phi]{\bf e_z}+i\text{exp}[i(l_{sy}-l_{0y}+l_{0z})\phi]{\bf e_z}),
\end{aligned}
\end{equation}
\begin{equation}
\begin{aligned}
\label{Eq:A0'_2}
{\bf A'_0}\propto
&({\bf A'_{00}\cdotp\frac{{\bf A_{s0}^*}}{|{\bf A_{s0}}|}})\frac{{\bf A_s}}{|{\bf A_{s0}}|}\propto
\\&
(\text{exp}[i(l_{0y})\phi]{\bf e_y}+\text{exp}[i(l_{0z}-l_{sz}+l_{sy})\phi]{\bf e_y}+
\\&
i\text{exp}[i(l_{0z})\phi]{\bf e_z}+i\text{exp}[i(l_{0y}-l_{sy}+l_{sz})\phi]{\bf e_z}).
\end{aligned}
\end{equation}
Therefore, Langmuir waves with OAM of $l_{0y}-l_{sy}$ and $l_{0z}-l_{sz}$ are generated. Also, seed pulses with new OAM of $l_{sz}-l_{0z}+l_{0y}$ in y polarisation and $l_{sy}-l_{0y}+l_{0z}$ in z polarisation, and pump pulses with new OAM of $l_{0z}-l_{sz}+l_{sy}$ in y polarisation and $l_{0y}-l_{sy}+l_{sz}$ in z polarisation are generated.

\begin{figure*}[!tp]
	\includegraphics[width=1.5\columnwidth]{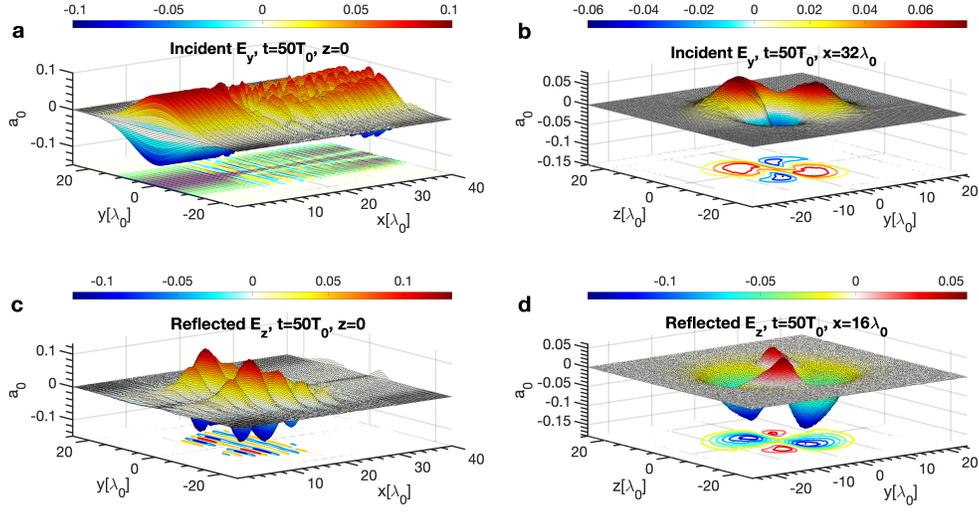}
	
	\caption{\label{Fig:Figure1}(Color online) Generation of the new pump (incident $E_y$) and seed (reflected $E_z$) beams with new OAM states. (a) The electric field in the pump incident direction ($+x$ direction) and the $y$ polarisation direction, labelled \textquoteleft Incident $E_y$'  which is defined as $1/2\cdot(E_y+c\cdot B_z)$. (b) The corresponding incident $E_y$ amplitude distribution in $yz$ plane at the position of $x=32\lambda_0$. (c) The electric field in the reflected direction ($-x$ direction) and z polarisation direction, labelled \textquoteleft Reflected $E_z$' which is defined as $1/2\cdot(E_z+c\cdot B_y)$. (d) The corresponding reflected $E_z$ amplitude distribution in $yz$ plane at the position of $x=16\lambda_0$.
		The OAM of the initial pump pulse is $l_{0y}=0$ and $l_{0z}=1$. The OAM of the initial seed pulse is $l_{sy}=1$ and $l_{sz}=0$. The time is $t=50T_0$. }
\end{figure*}

\begin{figure}[!tp]
	\includegraphics[width=1\columnwidth]{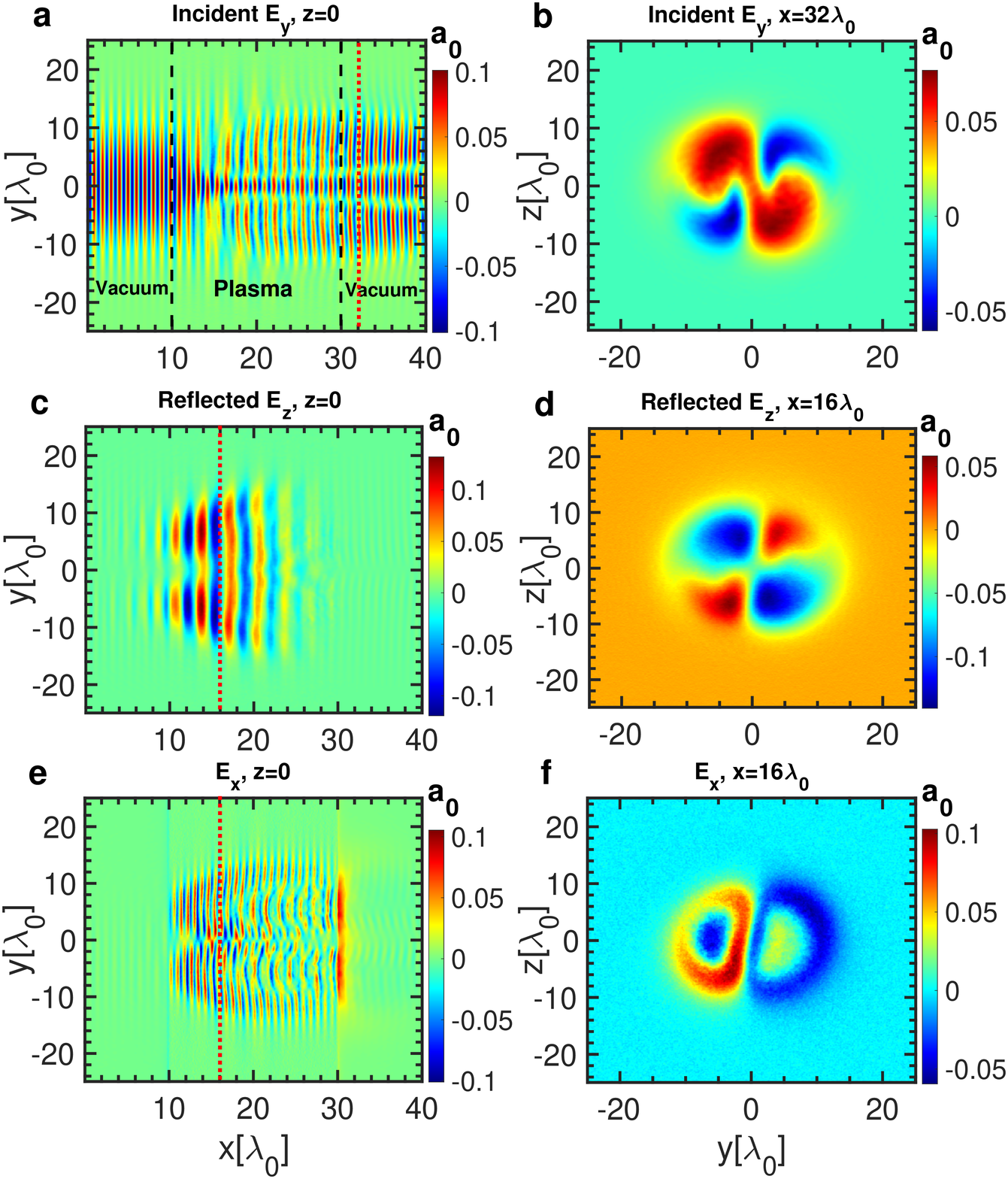}
	
	\caption{\label{Fig:Figure2}(Color online) Generation of the pump, seed and Langmuir wave with new OAM. The incident $E_y$ distribution in (a) $xy$ plane at the fixed position of $z=0$ and (b) $yz$ plane at the fixed position of $x=32\lambda_0$ which is labelled with the red dotted line in panel (a). The reflected $E_z$ distribution in (c) $xy$ plane at the fixed position of $z=0$ and (d) $yz$ plane at $x=16\lambda_0$ which is labelled with the red dotted line in panel (c). The longitudinal electric field $E_x$ distribution in (e) $xy$ plane at $z=0$ and (f) $yz$ plane at $x=16\lambda_0$. The initial conditions are the same as those in Figure \ref{Fig:Figure1}. The time is $t=50T_0$.}
\end{figure}

One-dimensional (1D)  and three-dimensional (3D) versions of fully relativistic kinetic Particle-in-Cell (PIC) code EPOCH \cite{Arber_2015PPCF,Feng_2020NF,Feng_2020SR} were used to investigate the Raman amplification of OAM beams. 
The electron temperature was $T_e=\SI{2.5}{keV}$ and electron density was $n_e=0.2n_c$, where $n_c$ was the critical density of the incident pump light. The ion temperature was $T_i=T_e/3$. For the backward Raman amplification cases, the spatial domain along x direction was set at [0, $40\lambda_0$] discretized with $N_x=1000$ spatial grid points and spatial step $dx=0.04\lambda_0$. The particles occupied the center simulation space of the box [$10\lambda_0$, $30\lambda_0$]  with two vacuum layers on either side of the plasma slab. The total simulation time was $t_{end}=80T_0$, where $T_0=\SI{1.17}{fs}$ was the period of $3\omega$ pump light (\SI{351}{nm}). The pump and seed pulse intensities were $I_0=1\times10^{17}\, \SI{}{W/cm^2}$ for both 1D and 3D simulations. The wavelength of pump pulse was $\lambda_0=\SI{351}{nm}$ and that of seed pulse was $\lambda_s=1.885\lambda_0$ for the backward Raman cases and $\lambda_s=1.818\lambda_0$ for the forward Raman cases. The time profile of the seed pulse was a Gaussian envelop $\text{exp}[(t-t_p)/\tau_0]$ with characteristic time $\tau_0=10T_0$ and peak time $t_{p}=20T_0$, and the pump met the seed at the right boundary of the plasma.

In 3D simulations, the simulation parameters were the same as those in the 1D simulations, but the spatial domains in y and z directions were [-$25\lambda_0$, $25\lambda_0$] and [-$25\lambda_0$, $25\lambda_0$] discretized with $N_y=500$ and $N_z=500$ spatial grid points. Open boundary conditions for the laser and particles were used in the $x$-direction for both 1D and 3D simulations, and periodic boundaries were used in $y$ and $z$ directions for the 3D simulations. There were 100 electrons and 100 ions per cell in the 1D simulations and 10 electrons/ions per cell in the 3D simulations.

Figures \ref{Fig:Figure1} and \ref{Fig:Figure2} show the generation of the new pump and seed with new OAM when the initial OAM conditions of pump and seed are $l_{0y}=0, l_{0z}=1$ and $l_{sy}=1, l_{sz}=0$. In y and z polarisation directions, pump couples with seed to generate Langmuir wave with OAM $l_{py}=l_{0y}-l_{sy}=0-1=-1$ and $l_{pz}=l_{0z}-l_{sz}=1-0=1$ which is clarified by Eq. (\ref{Eq:Density2}) and verified by Fig. \ref{Fig:Figure2}f. As shown in Fig. \ref{Fig:Figure2}f, the Langmuir wave with OAM of $l_{py}=-1$ is inside that with OAM of $l_{pz}=1$ due to the different spot size of initial OAM beam and Gaussian beam (with no OAM), although the beam waist at the focal plane is $w_0=10\lambda_0$ for both the pump and seed pulses. Note that $l_{py}$ and $l_{pz}$ are just to distinguish what polarisation direction the Langmuir wave is generated from, in fact both Langmuir waves with OAM of $l_{py}$ and $l_{pz}$ are in $x$ direction. 

Because the Langmuir wave is a longitudinal wave along the x direction, every transverse electromagnetic wave can couple with the Langmuir waves to generate new OAM beams. 

Firstly, the seed in the $y$ polarisation direction couples with the Langmuir wave with OAM $l_{pz}=1$ to generate a new pump with new OAM $l'_{0y}=l_{pz}+l_{sy}=l_{0z}-l_{sz}+l_{sy}=2$ which is mixed with the initial depleted pump with OAM $l_{0y}=0$ and demonstrated clearly in Figs. \ref{Fig:Figure1}b and \ref{Fig:Figure2}b. The seed in the $z$ polarisation direction couples with the Langmuir wave with OAM $l_{py}=-1$ to generate a new pump beam with new OAM $l'_{0z}=l_{py}+l_{sz}=l_{0y}-l_{sy}+l_{sz}=-1$ which is inside the initial depleted pump beam with OAM $l_{0z}=1$ (this mode is exactly observed but not shown here).

Secondly, the pump beam in the $z$ polarisation direction couples with the Langmuir wave with OAM $l_{py}=-1$ to generate a new seed with new OAM $l'_{sz}=l_{0z}-l_{py}=l_{0z}-(l_{0y}-l_{sy})=2$ which is mixed with the initial amplified seed with OAM $l_{sz}=0$ and demonstrated clearly in Figs. \ref{Fig:Figure1}d and \ref{Fig:Figure2}d. In addition, the pump beam in $y$ polarisation direction couples with the Langmuir wave with OAM $l_{pz}=1$ to generate a new seed with new OAM $l'_{sy}=l_{0y}-l_{pz}=l_{0y}-(l_{0z}-l_{sz})=-1$ which cannot be distinguished from the initial amplified seed with OAM $l_{sy}=1$ (not shown here). 

The other parameters are the same as Fig. \ref{Fig:Figure1}, but electric field amplitude of one polarisation of the pump or seed beam is set to be zero. These similar simulations are conducted but the results are not shown here. When linear-polarisation (LP) pump beam with $A_{0,y}=0$ is used and other parameters of pump and seed pulses are the same as those in Fig. \ref{Fig:Figure1}, i.e. LP pump with OAM $l_{0z}=1$ and CP seed with $l_{sy}=1$ and $l_{sz}=0$, new pump with OAM $l'_{0y}=l_{0z}-l_{sz}+l_{sy}=2$ is generated in $y$ polarisation direction. When LP pump with OAM $l_{0y}=0$ and CP seed with $l_{sy}=1$ and $l_{sz}=0$, new pump with OAM $l'_{0z}=l_{0y}-l_{sy}+l_{sz}=-1$ is generated in $z$ polarisation. When CP pump with OAM $l_{0y}=0$ and $l_{0z}=1$ and LP seed with $l_{sz}=0$, new seed with OAM $l'_{sy}=l_{0y}-(l_{0z}-l_{sz})=-1$ is generated in $y$ polarisation. When CP pump with OAM $l_{0y}=0$ and $l_{0z}=1$ and LP seed with $l_{sy}=1$, new seed with OAM $l'_{sz}=l_{0z}-(l_{0y}-l_{sy})=2$ is generated in $z$ polarisation.
The new pump and seed pulses with new OAM are consistent to the theoretical prediction as shown in Eqs. (\ref{Eq:As_1}) and (\ref{Eq:A0'_1}). 

\begin{figure}[!tp]
	\includegraphics[width=1\columnwidth]{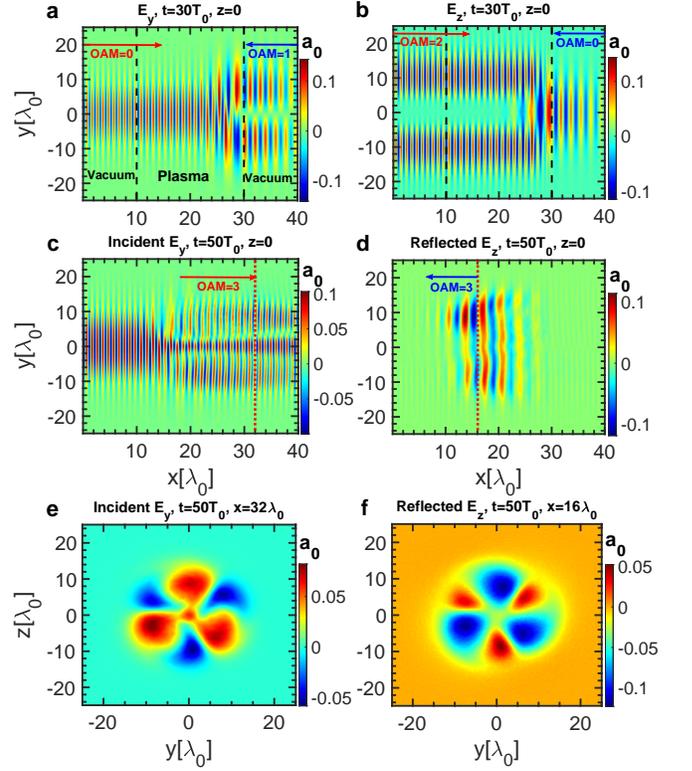}
	\caption{\label{Fig:Figure3}(Color online) Generation of the pump and seed pulses with new OAM. (a) $E_y$ and (b) $E_z$ distribution in $xy$ plane at $z=0$ and $t=30T_0$ when the pump and seed beams meet. (c) The incident $E_y$ and (d) the reflected $E_z$ distribution in $xy$ plane at $z=0$ and $t=50T_0$ when the pump and seed beams interact. Distribution in $yz$ plane of (e) the incident $E_y$ at $x=32\lambda_0$ and (f) the reflected $E_z$ at $x=16\lambda_0$.	The OAM of the initial pump pulse is $l_{0y}=0$ and $l_{0z}=2$. The OAM of the initial seed pulse is $l_{sy}=1$ and $l_{sz}=0$.}
\end{figure}

When the initial OAM conditions of the pump and seed pulses are $l_{0y}=0, l_{0z}=2$ and $l_{sy}=1, l_{sz}=0$, beams with higher-order OAM states are also generated, as shown in Fig. \ref{Fig:Figure3}. The pump beam with no OAM in $y$ polarisation direction interacts with the seed beam to generate a new pump beam with new OAM $l'_{0y}=3$ in this polarisation direction, as shown in Fig. \ref{Fig:Figure3}e. 
Because the seed beam with OAM $l_{sy}=1$ in $y$ polarisation direction couples with the Langmuir wave with OAM $l_{pz}=l_{0z}-l_{sz}=2-0=2$ to generate this new pump beam with higher OAM $l'_{0y}=l_{sy}+l_{pz}=3$. This new mode is clarified by the theoretical model of Eq. (\ref{Eq:A0'_1}).

In addition, the pump beam with OAM $l_{0z}=2$ in $z$ polarisation couples with the Langmuir wave with OAM $l_{py}=l_{0y}-l_{sy}=0-1=-1$ to generate new seed beam with a higher OAM state $l'_{sz}=l_{0z}-l_{py}=3$ in $z$ polarisation as shown in Fig. \ref{Fig:Figure3}f. These results give a way to produce both new long-duration pump and short-duration seed pulses with higher OAM by backward stimulated Raman scattering.

\begin{figure}[!tp]
	\includegraphics[width=1\columnwidth]{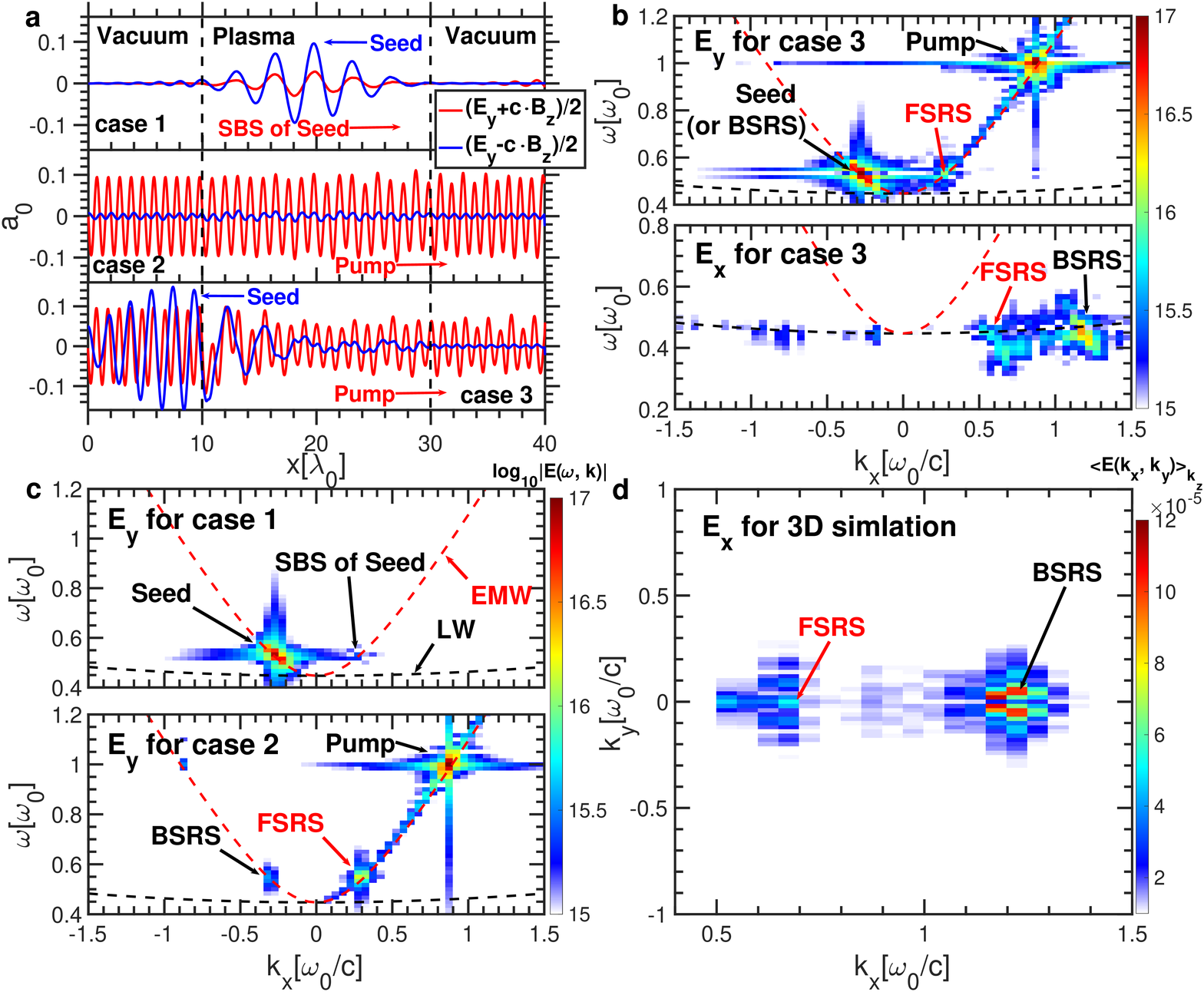}
	\caption{\label{Fig:Figure4}(Color online) The pump and seed propagation and corresponding spectra. (a) Only seed pulse in case 1 at $t=50T_0$, only pump pulse in case 2 at $t=50T_0$, and both pump and seed in case 3 at $t=60T_0$ are added in the system by 1D simulations. (b) Spectra of $E_y$ and $E_x$ for case 3 among the time range $[0, 60]T_0$ and spatial range $[10, 30]\lambda_0$. (c) Spectra of $E_y$ for case 1 and case 2 among $[0, 60]T_0$ and $[10, 30]\lambda_0$. (d) Wave-number spectrum of $E_x$ for the 3D simulations in Figure \ref{Fig:Figure1} at the fixed time $t=50T_0$ and among the local region $[10, 30]\lambda_0$. }
\end{figure}

Figures \ref{Fig:Figure4}a-\ref{Fig:Figure4}c show 1D simulation results from linearly polarised pump and seed beams, with the other conditions the same as those in the 3D simulations. In case 1, only seed pulse is added into the system. The amplitude of seed beam remains near to its initial amplitude, even while weak stimulated Brillouin scattering (SBS) of the seed pulse occurs as shown in the top panels of Figs. \ref{Fig:Figure4}(a) and \ref{Fig:Figure4}(c). In case 2, only the pump beam is added into the system. There is nearly no pump depletion in such short-scale plasma before the pump and seed pulses meet. Although signals of backward-SRS and forward-SRS from the pump beam occur as shown in the bottom panel of Fig. \ref{Fig:Figure4}(c), the amplitude of these scattering lights is very weak, as shown in the middle panel of Fig. \ref{Fig:Figure4}(a). In case 3, where both the pump and seed beams are added into the system, the seed is amplified to $a_s=0.15$ (or $I_s=2.5\times10^{17}$ $W/cm^2$) at $t=60T_0$ from the initial amplitude $a_{s0}=0.095$ (or $I_{s0}=1\times10^{17}$ $W/cm^2$). The amplitude of the amplified seed pulse at $t=60T_0$ in 1D simulations is nearly the same as that in 3D simulations with the CP-LG seed pulse. Because in our 3D simulations, the Raleigh length $x_R=\pi w_0^2/\lambda_0=314\lambda_0\gg L_x=40\lambda_0$, the paraxial approximation is satisfied. Thus, these results verify that the three wave coupling equations (\ref{Eq:LaserPropagation_pump_envelop1})-(\ref{Eq:Density_envelop1}) are independent of the laser transverse profiles, which are the same for both the linearly polarised lasers in the 1D simulations and CP-LG lasers in the 3D simulations.

In case 3, the dominant instability is backward-SRS, as shown in Fig. \ref{Fig:Figure4}b, because the seed is with the matched frequency of backward-SRS and also the growth rate of backward-SRS is larger than that of forward-SRS. However, the forward-SRS spectra also occur in the interaction process for both the 1D (Fig. \ref{Fig:Figure4}b) and the 3D (\ref{Fig:Figure4}d) simulations. This illustrates that the forward-SRS can also be taken as a potential mechanism to amplify the seed and generate new OAM beams, which will be shown later. 

In our simulations, i.e. when $n_e=0.2n_c, T_e=\SI{2.5}{keV}$ and $\lambda_0=\SI{351}{nm}$, the growth rate of backward-SRS is very large, which is much higher than that in low density regime. Therefore, the seed pulse is amplified and enters into the nonlinear regime quickly. On one hand, the seed pulse is able to be amplified to high intensity in very short scale plasma. On the other hand, the energy transfer efficiency of Raman amplification in this high electron density regime is much higher than that of usual Raman amplification in very low density regime. 

We have done variations of 1D and 3D simulations with many other parameters.
The main results in the optimal parameter regimes are as follows:
(i) The seed has been amplified from initial intensity $I_{s0}=1\times 10^{16}$ $W/cm^2$ (FWHM duration $\tau_{s0}=16.7T_0$) to $I_s=6.9\times 10^{16}$ $W/cm^2$ ($\tau_{s}\simeq15T_0$) by the pump with intensity $I_0=1\times10^{16}$ $W/cm^2$ in a plasma with scale of $L_p=75\lambda_0$. The energy transfer efficiency is $r\simeq61\%$. (ii) The seed has been amplified from initial intensity $I_{s0}=4\times 10^{16}$ $W/cm^2$ ($\tau_{s0}=16.7T_0$) to $I_s=1.1\times 10^{16}$ $W/cm^2$ ($\tau_{s}\simeq16.7T_0$) by the pump with intensity $I_0=4\times10^{16}$ $W/cm^2$ in a plasma with scale of $L_p=40\lambda_0$. The energy transfer efficiency is $r\simeq40\%$. (iii) As shown in the 1D simulation in Fig. \ref{Fig:Figure4} and the 3D simulations in Fig. \ref{Fig:Figure1}, the seed pulse has been amplified from initial intensity $I_{s0}=1\times 10^{17}$ $W/cm^2$ ($\tau_{s0}=16.7T_0$) to $I_s=2.5\times 10^{17}$ $W/cm^2$ ($\tau_{s}\simeq15T_0$) by the pump beam with intensity $I_0=1\times10^{17}$ $W/cm^2$ in a plasma with scale of $L_p=20\lambda_0$. The energy transfer efficiency is $r\simeq56\%$, which is much higher than that of the usual Raman amplification in very low density regimes \cite{Ren_2007NP} and confirm that we are operating in the non-linear regime, identified by Sadler \textit{et al.} \cite{Sadler_2017PRE} and Trines \textit{et al} \cite{Trines_2020SR}. The seed pulse envelop quality in the above processes (i)-(iii) are very good without filamentation, wave-breaking and strong parasitic stimulated Raman/Brillouin scattering, because very short scale plasma is taken as the medium to prevent these harmful instabilities from developing before or during the Raman amplification process. These results give a promising path to amplify the seed by multi-stage stimulated Raman scattering more efficiently in high electron density regime.

\begin{figure}[!tp]
	\includegraphics[width=1\columnwidth]{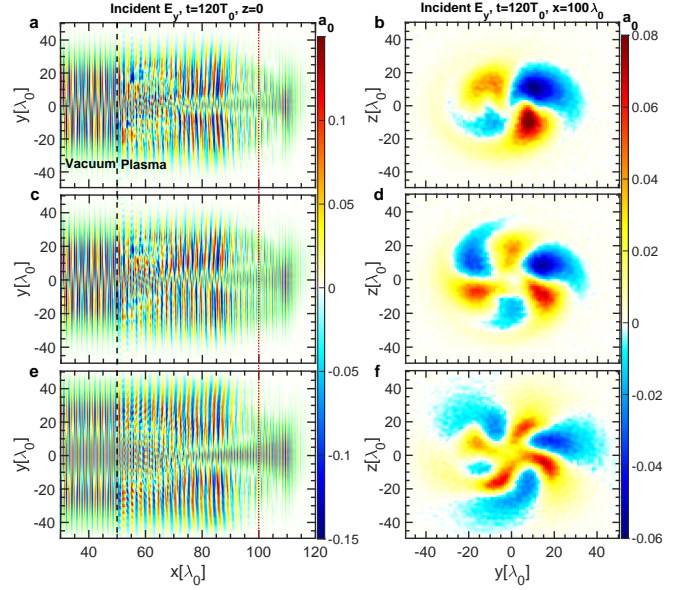}
	
	\caption{\label{Fig:Figure5}(Color online) Pump with new OAM generation by forward stimulated Raman scattering. (a), (c), (e) The incident $E_y$ in $xy$ plane at $t=120T_0$ and $z=0$. (b), (d), (f) The corresponding incident $E_y$ distribution in $yz$ plane at $t=120T_0$ and $x=100\lambda_0$ which are labelled with red dotted lines in panels (a), (c), and (e).  The pump and seed are propagating into the system at the same direction. The OAM of the pump and seed in panels (a, b) are $l_{0y}=0, l_{0z}=1$ and $l_{sy}=1, l_{sz}=0$. The OAM of the pump and seed in panels (c, d) are $l_{0y}=0, l_{0z}=2$ and $l_{sy}=1, l_{sz}=0$. The OAM of the pump and seed in panels (e, f) are $l_{0y}=0, l_{0z}=1$ and $l_{sy}=3, l_{sz}=0$.    }
\end{figure}

Figure \ref{Fig:Figure5} shows the generation of the pump beam with new OAM states by the forward-SRS mechanism. The pump and seed pulses propagate into the system at the same time. The duration of the seed is the same as that of the pump beam, and the seed wavelength is $\lambda_s=1.818\lambda_0$, where $\lambda_0$ is the pump wavelength. The spatial domain along x direction is $[0, 200\lambda_0]$. And the particles occupy the center of the box $[50, 150]\lambda_0$ with two vacuum layers on either side of the plasma slab. For forward-SRS when $n_e=0.2n_c, T_e=\SI{2.5}{keV}$, group velocity of forward-SRS is $v_{gs}^F/c=ck_s/\omega_s=0.32/0.547=0.585$, and group velocity of pump is $v_{g0}/c=ck_0/\omega_0=\sqrt{1-n_e/n_c}=0.894$. At $t=120T_0$, the pump and seed leading edges reach positions of $x_0=113\lambda_0$ and $x_s=91\lambda_0$. Therefore, at the position of $x=100\lambda_0$ as shown in Fig. \ref{Fig:Figure5}, only pump can reach here and the signals at this position are pump signals. As shown in Figs. \ref{Fig:Figure5}b, \ref{Fig:Figure5}d and \ref{Fig:Figure5}f, new pumps with higher OAM $l'_{0y}=l_{0z}-l_{sz}+l_{sy}$ are generated by the forward-SRS mechanism. When $l_{0y}=0, l_{0z}=1$ and $l_{sy}=1, l_{sz}=0$, pump with new OAM $l'_{0y}=2$ is generated as shown in Fig. \ref{Fig:Figure5}b. When $l_{0y}=0, l_{0z}=2$ and $l_{sy}=1, l_{sz}=0$, pump with new OAM $l'_{0y}=3$ is generated as shown in Fig. \ref{Fig:Figure5}d. When $l_{0y}=0, l_{0z}=1$ and $l_{sy}=3, l_{sz}=0$, pump with new OAM $l'_{0y}=4$ is generated as shown in Fig. \ref{Fig:Figure5}f. We have also done the simulations of short duration of seed propagating at the same direction of the pump, the new OAM pumps are also generated, which is similar with the phenomenon of long duration seed cases. These results give a path to generate new pump with higher OAM by forward-SRS mechanisms.

In conclusions, this article gives a promising way to generate long duration pump pulses with new OAM by both backward and forward stimulated Raman scattering for the first time. Also, the short duration seed with higher OAM is generated and efficiently amplified in very short plasmas, and the amplification efficiency shown in this article is much higher than previous schemes in low density plasmas. These results show the potential to efficiently amplify the seed to high intensity by multi-stage backward-SRS in high electron density regime.
The experiments related to the results presented in this article are possible to be conducted in current laser facilities, which will be conducted in the future.

This article shows the potential applications in probing the OAM of continuous media using the stimulated Raman scattering mechanism. Stimulated Brillouin scattering, the ion acoustic relative of SRS, is also predicted to generate and amplify new OAM beams. Conclusive evidence has been provided that the seed pulse couples with an OAM Langmuir wave to generate a new OAM pump beam (or pump can couple with OAM Langmuir wave to generate new OAM seed), thus the OAM information of laser can be taken as the important role to explore the OAM information of plasma and other continuous media by stimulated Raman/Brillouin scattering mechanisms. These results shown in this article will be of wide interests and have potential applications in high-energy-density physics \cite{Vieira_2018PRL,Ramy_2019PRL,Zhu_2016NC,Zhu_2018NJP}, nonlinear optics \cite{Li_2016NP}, and condensed matter physics \cite{Martinelli_2004PRA}.  

		We would like to acknowledge useful discussions with L. B. Ju, C. Z. Xiao and Q. Wang. The authors gratefully acknowledge the support of the  ARCHER2 UK National Supercomputing Service, all of the staff of the Central Laser Facility and the Scientific Computing Department's SCARF supercomputing facility at the UKRI-STFC Rutherford Appleton Laboratory. 
		This research was supported by the Oxford-ShanghaiTech collaboration, the UKRI-EPSRC funded e674 ARCHER2 project under grant number EP/R029148/1 and the National Natural Science Foundation of China (Grant Nos. 12005021).

\bibliography{FSRS_OAM}

\end{document}